\documentclass[10pt]{article}
\usepackage{amssymb}
\usepackage{amsmath}
\usepackage{graphicx}
\usepackage{fancyhdr}
\usepackage{cite}

\begin{document}

\newcommand{\threej}[6]{\left(\begin{array}{ccc}#1 & #2 & #3 \\ #4 & #5 & #6 \end{array}\right)}

\huge

\begin{center}
A new sum rule for Clebsch-Gordan coefficients using generalized characters of irreducible representations of the rotation group
\end{center}

\vspace{0.5cm}

\large

\begin{center}
Jean-Christophe Pain\footnote{jean-christophe.pain@cea.fr}
\end{center}

\normalsize

\begin{center}
CEA, DAM, DIF, F-91297 Arpajon, France
\end{center}

\vspace{0.5cm}


\begin{abstract}
We present a new sum rule for Clebsch-Gordan coefficients using generalized characters of irreducible representations of the rotation group. The identity is obtained from an integral involving Gegenbauer ultraspherical polynomials. A similar procedure can be applied for other types of integrals of such polynomials, and may therefore lead to the derivation of further new relations.
\end{abstract}

\section{Introduction}

Several special relations (identities, sum rules) involving Clebsch-Gordan coefficients or Wigner $3jm$ symbols have been discovered in connection with atomic, molecular and nuclear spectroscopy (see the non-exhaustive list of references \cite{dunlap75,morgan76,rashid76,vandenberghe76,morgan77,demeyer78,klarsfeld78,din81,askey82,elbaz85,brudno85,kancerevicius90,ginocchio91,minnaert94,casini97,pain12,rowe97,speziale17,ibort17}). Applications concern for instance the hydrogen molecular ion \cite{dunlap75}, the non-relativistic helium atom \cite{morgan76,rashid76,vandenberghe76,morgan77,demeyer78}, the high-order radiative transitions in hydrogenic ions \cite{klarsfeld78}, the stability properties of some special classical solutions of the $O(n)$ non-linear $\sigma$-model in two dimensions \cite{din81,askey82}, the non-trivial zeros of $3jm$ and $3nj$ coefficients \cite{brudno85}, the pion double charge exchange cross-sections in the nuclear shell model \cite{ginocchio91}, the Stark effect of hydrogenic systems \cite{casini97}, or the statistical modeling of anomalous Zeeman effect \cite{pain12}. Some of these identities are sometimes referred to as ``unusual sum rules'', in the sense that they can not be reduced to orthogonality relations, or that they do not include the $(2j+1)$ weighting factor in a summation over angular momentum $j$ \cite{elbaz85}. Sum rules can be of great interest for checking numerical calculations involving Clebsch-Gordan or Wigner $3jm$ symbols. The reference book of Varshalovich, Moskalev and Khersonskii \cite{varshalovich88} presents (section 8.7.7, p. 262) three unusual sum rules involving $3jm$ symbols \cite{dunlap75,morgan76,din81} and only one consisting of a summation over projection of angular momentum $j$ ($-j\leq m\leq j$), a relation obtained by Dunlap and Judd \cite{dunlap75}. In this work we present, using the connections between generalized characters of irreducible representations and a particular integral involving a product of ultraspherical Gegenbauer polynomials, a new sum rule for Clebsch-Gordan coefficients. To the best of our knowledge, such a result has not been reported before. The sum rule is also given in terms of $3jm$ symbols. 

\section{Derivation of the new sum rule}

The generalized character (of order $k$) of the irreducible representation of rank $j$ of the rotation group is defined as

\begin{equation}
\chi_k^j(\omega)=\sqrt{\frac{(2j+1)(2j-k)!}{(2j+k+1)!}}\sin^k\left(\frac{\omega}{2}\right)\left(\frac{d}{d\cos\left(\frac{\omega}{2}\right)}\right)^k\chi^j(\omega),
\end{equation}

\noindent where $k$ is integer ($0\leq k\leq 2j$) and $\chi^j$ is the character of the irreducible representation of rank $j$. The generalized character can be written \cite{talman68,varshalovich88} as

\begin{equation}\label{chi1}
\chi_k^j(\omega)=i^k\sum_me^{-im\omega}\langle jmk0|jm\rangle,
\end{equation}

\noindent where $\langle j_1m_1j_2m_2|j_3m_3\rangle$ is the usual Clebsch-Gordan coefficient \cite{cowan81}. $\chi_k^j(\omega)$ can also be expressed in terms of Gegenbauer (or ultraspherical) \cite{abramowitz64} polynomials $C_n^{(\alpha)}$:

\begin{equation}\label{chi2}
\chi_k^j(\omega)=(2k)!!\sqrt{\frac{(2j+1)(2j-k)!}{(2j+k+1)!}}\sin^k\left(\frac{\omega}{2}\right)C_{2j-k}^{(k+1)}\left(\cos\left(\frac{\omega}{2}\right)\right),
\end{equation}

\noindent where $(2k)!!=2k\times(2k-2)\times(2k-4)\times\cdots\times 4\times 2=2^kk!$. Gazeau and Kibler \cite{gazeau80} obtained a sum rule on $3jm$ symbols using Bander-Itzykhson polynomials. Their expression can be obtained from the orthogonality relation for generalized characters

\begin{equation}
\int_0^{2\pi}\chi_k^j(\omega)\chi_k^{j'}(\omega)\sin^2\left(\frac{\omega}{2}\right)d\omega=\pi\delta_{j,j'},
\end{equation}

\noindent $\delta_{a,b}$ being Kronecker's symbol. The latter equation is equivalent to the orthogonality relation for Gegenbauer poynomials:

\begin{equation}\label{norm}
\int_{-1}^1C_n^{(\alpha)}(x)C_r^{(\alpha)}(x)\left(1-x^2\right)^{\alpha-1/2}dx=\frac{\pi 2^{1-2\alpha}\Gamma(n+2\alpha)}{n!(n+\alpha)\left[\Gamma(\alpha)\right]^2}\delta_{n,r},
\end{equation}

\noindent where $\Gamma$ represents the Gamma function. Integrals involving products of Gegenbauer polynomials \cite{gradshteyn80} is a powerful tool for finding identities which may be of great interest for atomic, molecular and nuclear physics \cite{rahman85,vanhaeringen86}. Laursen and Mita obtained, for $\alpha\geq 0$, the following expression \cite{laursen81}:

\begin{eqnarray}\label{normn}
\int_{-1}^1\left[C_n^{(\alpha)}(x)\right]^2\left(1-x^2\right)^{\alpha+\mu-1/2}dx&=&\frac{\pi\Gamma(2\alpha+n)\Gamma(\alpha+\mu+1/2)\Gamma(\alpha-\mu+n)}{2^{2\alpha-1}(\alpha+\mu+n)n!\Gamma(\alpha)\Gamma(\alpha-\mu)\Gamma(\alpha+1/2)\Gamma(\alpha+\mu+n)}\nonumber\\
& &\times~_4F_3\left[\begin{array}{l}
-n,2\alpha+n,1/2,-\mu\\
\alpha+1/2,\alpha-\mu,1
\end{array};1\right],
\end{eqnarray}

\noindent where $~_4F_3$ represents the hypergeometric function. The latter expression becomes, in the particular case $\mu$=1,

\begin{equation}\label{normn2}
\int_{-1}^1\left[C_n^{(\alpha)}(x)\right]^2\left(1-x^2\right)^{\alpha+1/2}dx=\frac{\pi\Gamma(2\alpha+n)\left[(\alpha-1)(\alpha+1/2)+n(\alpha+n/2)\right]}{2^{2\alpha-1}(\alpha+n-1)(\alpha+n)(\alpha+n+1)n!\left[\Gamma(\alpha)\right]^2}.
\end{equation}

\noindent Setting $\eta=\omega/2$, one gets, from Eqs. (\ref{chi1}) and (\ref{chi2}):

\begin{equation}\label{chi1p}
\chi_k^j(\eta)\chi_k^{j*}(\eta)=\sum_{m,m'}e^{-2i(m-m')\eta}\langle jmk0|jm\rangle\langle jm'k0|jm'\rangle
\end{equation}

\noindent and

\begin{equation}\label{chi2p}
\chi_k^j(\eta)\chi_k^{j*}(\eta)=(2^kk!)^2\frac{(2j+1)(2j-k)!}{(2j+k+1)!}\left[C_{2j-k}^{(k+1)}\left(\cos\eta\right)\right]^2\sin^{2k}\eta
\end{equation}

\noindent respectively. Equating the two right-hand sides of Eqs. (\ref{chi1p}) and (\ref{chi2p}) yields, after multiplication by $\sin^4\eta$ and integration from $\eta$=0 to $\eta$=$\pi$:

\begin{eqnarray}
& &\sum_{m,m'}\langle jmk0|jm\rangle\langle jm'k0|jm'\rangle\int_0^{\pi}e^{-2i(m-m')\eta}\sin^4\eta d\eta\nonumber\\
& &\;\;\;\;\;\;\;\;=\left(2^kk!\right)^2\frac{(2j+1)(2j-k)!}{(2j+k+1)!}\int_0^{\pi}\left[C_{2j-k}^{(k+1)}\left(\cos\eta\right)\right]^2\sin^{2k+4}\eta d\eta.
\end{eqnarray}

\noindent Setting $x=\cos\eta$, $\alpha=k+1$ and $n=2j-k$, Eq. (\ref{normn2}) becomes

\begin{equation}
\int_0^{\pi}\left[C_{2j-k}^{(k+1)}\left(\cos\eta\right)\right]^2\sin^{2k+4}\eta d\eta=\frac{\pi}{16}\frac{(2j+k+1)!}{(2j-k)!(2j+1)j(j+1)}\frac{\left[k(k+1)+4j(j+1)\right]}{\left(2^kk!\right)^2}
\end{equation}

\noindent and thus

\begin{equation}
\sum_{m,m'}\langle jmk0|jm\rangle\langle jm'k0|jm'\rangle\int_0^{\pi}e^{-2i(m-m')\eta}\sin^4\eta d\eta=\frac{\pi\left[k(k+1+4j(j+1)\right]}{16j(j+1)},
\end{equation}

\noindent with

\begin{eqnarray}
\int_0^{\pi}e^{-2i(m-m')\eta}\sin^4\eta d\eta &=&\frac{3}{2}\frac{(-1)^{m'-m}\sin\left[\left(m'-m\right)\pi\right]}{\left(m'-m\right)\left[1-\left(m'-m\right)^2\right]\left[4-\left(m'-m\right)^2\right]},\nonumber\\
\end{eqnarray}

\noindent yielding to the new sum rule for Clebsch-Gordan coefficients

\begin{eqnarray}\label{cg}
& &\sum_{m,m'}\frac{(-1)^{m'-m}\sin\left[\left(m'-m\right)\pi\right]}{\left(m'-m\right)\pi\left[1-\left(m'-m\right)^2\right]\left[4-\left(m'-m\right)^2\right]}\langle jmk0|jm\rangle\langle jm'k0|jm'\rangle\nonumber\\
& &\;\;\;\;\;\;\;\;=\frac{\left[k(k+1)+4j(j+1)\right]}{24j(j+1)}\delta(jkj),
\end{eqnarray}

\noindent where the notation $\delta(abc)$ (see Ref. \cite{edmonds57}) means that $a$, $b$ and $c$ satisfy triangular relations. Using Euler's sine product formula

\begin{equation}
\Gamma(x)\Gamma(1-x)=\frac{\pi}{\sin(\pi x)},
\end{equation}

\noindent Eq. (\ref{cg}) can also be put in the form

\begin{eqnarray}\label{cga}
& &\sum_{m,m'}\frac{(-1)^{m'-m}}{\Gamma\left(3+m-m'\right)\Gamma\left(3-m+m'\right)}\langle jmk0|jm\rangle\langle jm'k0|jm'\rangle\nonumber\\
& &\;\;\;\;\;\;\;\;=\frac{\left[k(k+1)+4j(j+1)\right]}{24j(j+1)}\delta(jkj).
\end{eqnarray}

\noindent Using the relation between Clebsch-Gordan coefficients and $3j$ symbols \cite{varshalovich88}:

\begin{equation}
\langle j_1m_1j_2m_2|j_3m_3\rangle=(-1)^{j_1-j_2+m_3}\sqrt{2j_3+1}\threej{j_1}{j_2}{j_3}{m_1}{m_2}{-m_3},
\end{equation}

\noindent as well as the symmetry property

\begin{equation}
\threej{j_3}{j_2}{j_1}{-m_3}{m_2}{m_1}=(-1)^{j_1+j_2+j_3}\threej{j_1}{j_2}{j_3}{m_1}{m_2}{-m_3},
\end{equation}

\noindent we find that Eq. (\ref{cg}) becomes 

\begin{eqnarray}\label{res3j}
& &\sum_{m,m'}\frac{\sin\left[\left(m'-m\right)\pi\right]}{\left(m'-m\right)\pi\left[1-\left(m'-m\right)^2\right]\left[4-\left(m'-m\right)^2\right]}\threej{j}{k}{j}{-m}{0}{m}\threej{j}{k}{j}{-m'}{0}{m'}\nonumber\\
& &\;\;\;\;\;\;\;\;=\frac{\left[k(k+1)+4j(j+1)\right]}{24j(j+1)(2j+1)}\delta(jkj).
\end{eqnarray}

\noindent Eq. (\ref{res3j}) can also be expressed as

\begin{eqnarray}\label{ga}
& &\sum_{m,m'}\frac{1}{\Gamma\left(3+m-m'\right)\Gamma\left(3-m+m'\right)}\threej{j}{k}{j}{-m}{0}{m}\threej{j}{k}{j}{-m'}{0}{m'}\nonumber\\
& &\;\;\;\;\;\;\;\;=\frac{\left[k(k+1)+4j(j+1)\right]}{24j(j+1)(2j+1)}\delta(jkj).
\end{eqnarray}

\noindent The identities (\ref{cg}), (\ref{cga}), (\ref{res3j}) and (\ref{ga}) have been tested numerically.

\section{Conclusion}

A new sum rule for Clebsch-Gordan coefficients or $3j$ Wigner symbols was derived using generalized characters of irreducible representations of the rotation group. Beyond this simple but non-trivial result, which may be of interest for angular-momentum calculations, we hope that the technique presented here shall stimulate the generation of new identities.


\end{document}